# Error Rates Analysis of MIMO Space-Time Block Codes in Generalized Shadowed Fading Channels

Ehab Salahat, Ali Hakam, and Nazar Ali

Khalifa University, Department of Electrical and Computer Engineering, Abu Dhabi, UAE

*Abstract*—This paper introduces a new and unified bit error rates performance analysis of space-time block codes (STBC) deployed in wireless systems with spatial diversity in generalized shadowed fading and noise scenarios. Specifically, we derive a simple and a very accurate approximate expressions for the average error rates of coherent modulation schemes in generalized $\eta$-$\mu$ and $\kappa$-$\mu$ shadowed fading channels with multiple input multiple output (MIMO) systems. The noise in the network is assumed to be modeled using the additive white generalized Gaussian noise (AWGGN), which encompasses the classical Laplacian and the Gaussian noise environments as special cases. The derived results obviate the need to re-derive the error rates for MIMO STBC systems under many multipath fading and noise conditions while avoiding any special functions with high computational complexity. Published results from the literature, as well as numerical evaluations, corroborate the accuracy of our derived generalized expressions.

## I. Introduction

Next Generation mobile communication is seen to increase data transfer rates in bandwidth-limited wireless channels. Wireless propagation channels, however, suffer from time-varying impairments that degrade the transmission quality and the performance, e.g. multipath fading and noise [1] [2]. Space-time coding is a famous technique utilized to solve bandwidth limitations and its associated problems based on Orthogonal Frequency Division Multiplexing [3]. Compared with the other coded communications system, space-time coding can effectively improve the capacity of wireless transmission in relatively simple receiver structure, while getting higher coding gain without sacrificing the whole bandwidth [4]. Moreover, using multiple transmitter and/or receiver antennas in MIMO systems enhances the reliability of wireless transmission significantly, mainly due to the property of spatial diversity. Some of the known and well-studied receiver diversity techniques include maximum ratio combining (MRC), selection combining (SC), and equal gain combining (EGC). However, due to some drawbacks of using diversity at the receiver's side (e.g. increased size due to the inclusion of more antenna), transmitter diversity seems a smarter design option. It can obtain the same performance gain for the downlink while moving the required extra antennas to the transmitter (e.g. the base station).

Space-time block codes (STBCs), introduced by *Alamouti* in his seminal work [5], and extended by pioneer researchers such as Tarokh [6], are frequently coupled with MIMO systems, being considered as the standard approach to combat multipath fading, in addition to their ability to achieve full-diversity as well as enhancing the end-to-end network performance [7] [8].

The literature is rich with research work on STBCs in MIMO systems that allowed better comprehension of their anticipated performance and design trade-offs. For example, an excellent survey paper on signal identification on MIMO and STBC can be found in [9]. Another example is in [10] where the authors proposed a new orthogonal STBC scheme that can resolve the linear decoding problem while keeping the simplicity and high performance properties of the classical OSTBC, and achieve the query diversity for the MIMO backscatter RFID. More recent examples include [11] where the authors presented a novel approach to MIMO signal identification by considering the modulation type and the STBC classification tasks as a joint classification problem.

The fading channels in MIMO systems are generally assumed, essentially for mathematical convenience, to be independent and identically distributed Rayleigh, Rician or Nakagami-*m* fading channels. For instance, *Ghosh et al.* presented in [12] an error analysis of multiple rate STBC for MIMO network over Rician channels. Nakagami-*m* fading was also used in [13] for the presented MIMO system. However, the use of simplified fading models may be physically unjustified for many practical scenarios of interest. In addition, despite their undoubted usefulness, those simplified models and fading assumptions limit their wider applicability. Moreover, most of the published research on MIMO STBC assumes short-term fading only, ignoring signal shadowing effect (a.k.a. large-scale fading). To resolve this issue, some researchers have considered the generalized-*K* model. For instance, the authors in [14] have analyzed generalized-*K* fading with MIMO channels and STBC. The average channel capacity for the same fading as well as error rates were also derived in [15] and [16], where the results were derived using computationally intensive special functions, which are not easily amenable for further analytical manipulations. In addition, up to our knowledge, most of the published research treats the calculations assuming Gaussian noise environment, i.e. the additive white Gaussian noise (AWGN) model.

To circumscribe these issues, in this work we present a new unified performance study of STBC operating in the MIMO network. New simple and very accurate average bit error rates (ABER) analytical expressions are derived in the generalized $\eta$-$\mu$ and $\kappa$-$\mu$ shadowed fading channels. It is worth mentioning here that it was recently shown in [17] that the $\kappa$-$\mu$ shadowed model actually includes the $\eta$-$\mu$ model as a special case. The additive noise in the network is modeled using the AWGGN, which also encloses many important noise models as particular cases. The new result obviates the need to re-derive similar expressions for the special cases of the fading and noise scenarios. Numerical techniques as well as published results from the public technical literature corroborate our results.

The remaining sections of this work are structured as follows. Section II presents the mathematical derivations of the associated system model, fading models, and noise environment. Derivation of the unified error rates expressions is presented in section III while the simulation results are studied in section IV. Finally, the paper findings and conclusion are given in section V.

## II. MATHEMATICAL DERIVATIONS

### A. The $\eta - \mu$ Fading Models

The power probability density function (PDF) of the first and second formats of the generalized $\eta - \mu$ model (namely, the $\eta - \mu$ and $\lambda - \mu$ formats, respectively) is given [18] [19] by:

$$f_\gamma(\gamma) = \frac{2\sqrt{\pi}\mu^{\mu+0.5}h^\mu}{\Gamma(\mu)H^{\mu-0.5}\tilde{\gamma}^{\mu+0.5}}\gamma^{\mu-0.5}e^{-\left[\frac{2\mu h}{\tilde{\gamma}}\right]\gamma}I_{\mu-0.5}\left(\left[\frac{2\mu H}{\tilde{\gamma}}\right]\gamma\right), \quad (1)$$

where $h$ and $H$ are as given in Table I for the two formats, $I_v(\cdot)$ denotes the modified Bessel function of the first type and order $v$ [20], and the physical wireless fading parameters $\lambda$, $\eta$, and $\mu$ signify the correlation of the in-phase (I) and quadrature (Q) components of the fading signal, the unequal power of these components, and the total number of multipath clusters [18].

TABLE I: $h$ AND $H$ VALUES FOR THE $\eta - \mu$ AND $\lambda - \mu$ MODELS [2]

| Distribution | $h$ | $H$ |
|---|---|---|
| $\eta - \mu$ | $0.25\eta^{-1}(1+\eta)^2$ | $0.25\eta^{-1}(1-\eta^2)$ |
| $\lambda - \mu$ | $(1-\lambda^2)^{-1}$ | $\lambda(1-\lambda^2)^{-1}$ |

This fading model includes the Rician, the Hoyt (Nakagami-$q$), the Nakagami-$m$, the Rayleigh and the One-Sided Gaussian models as special cases. Table I in [2] and Table II in [21] discusses the relation between this model and its special cases.

### B. The $\kappa - \mu$ Shadowed Fading Model

The $\kappa-\mu$ shadowed fading distribution mainly relies on the generalization of the $\kappa-\mu$ model and its physical parameters as discussed in [22, 23]. The model assumes a non-homogeneous signal propagation environment, and that the multipath scatters have the same power but an arbitrary power for the dominant component. This is different from the $\kappa-\mu$ distribution that adopts a deterministic dominant component assumption for every cluster, that is, the shadowed $\kappa-\mu$ model assumes that the power of the dominant components of each cluster changes arbitrarily due to the signal shadowing effect. Intuitively, since the $\kappa-\mu$ distribution encompasses the Rician model as a particular case [18], an intuitive generalization of the $\kappa-\mu$ fading model can be obtained by shadowing it with the same multipath/shadowing scheme used in the Rician shadowed model [24]. Assuming that multipath shadowing follows Nakagami-$m$, then the probability density function of the $\kappa-\mu$ shadowed fading is given as [24]:

$$f_\gamma(\gamma) = \frac{\mu^\mu m^m (1+\kappa)^\mu}{\Gamma(\mu)(\mu\kappa+m)^m \tilde{\gamma}^\mu}\gamma^{\mu-1}e^{-\left[\frac{\mu(1+\kappa)}{\tilde{\gamma}}\right]\gamma} {}_1F_1\left(m;\mu;\left[\frac{\mu^2\kappa(1+\kappa)}{[\mu\kappa+m]\tilde{\gamma}}\right]\gamma\right), \quad (2)$$

where $\mu$ is defined as before, whereas $\kappa$ and $m$ account for the ratio between the total power of the dominant components and the total power of the scattered waves and the shadowing parameter, respectively [18] [24], and ${}_1F_1(\cdot)$ denotes the hypergeometric function [20]. This distribution includes the $\kappa-\mu$, Nakagami-$m$, Rician shadowed, Rician, Rayleigh, one-sided Gaussian and the $\eta$-$\mu$ as special cases. These models can be achieved from (2) as shown in and summarized in Table II.

TABLE II: $\kappa-\mu$ SHADOWED FADING PARTICULAR CASES [17].

| Fading Distribution | $\mu$ | $\kappa$ | $m$ |
|---|---|---|---|
| $\kappa - \mu$ | $\mu$ | $\kappa$ | $m \to \infty$ |
| $\eta - \mu$ | $2\mu$ | $0.5\eta^{-1}(1-\eta)$ | $m = \mu$ |
| Rician shadowed | $\mu = 1$ | $\kappa = K$ | $m = m$ |
| Nakagami-$q$ (Hoyt) | 1 | $0.5q^{-2}(1-q^2)$ | $m = 0.5$ |
| Rician | $\mu = 1$ | $\kappa = K$ | $m \to \infty$ |
| Nakagami-$m$ | $\mu = m$ | $\kappa \to 0$ | $m \to \infty$ |
| Rayleigh | $\mu = 1$ | $\kappa \to 0$ | $m \to \infty$ |
| One-Sided Gaussian | $\mu = 0.5$ | $\kappa \to 0$ | $m \to \infty$ |

### C. Space-Time Block Coding Analysis

Following the same analysis and the assumptions given in [25], and assuming that the number of transmit antennas and receive antennas is given respectively by $N_t$ and $N_r$, it was shown in [25] that the received power PDF (uncorrelated antennas) for the $\eta - \mu$ generalized fading is given by [21]:

$$f_\gamma(\gamma) = \frac{2\sqrt{\pi}h^{\mu N_t N_r}}{\Gamma(\mu N_t N_r)}\left(\frac{\mu}{\tilde{\gamma}}\right)^{\mu N_t N_r + 0.5}\left(\frac{\gamma}{H}\right)^{\mu N_t N_r - 0.5} \times e^{-\left[\frac{2\mu h}{\tilde{\gamma}}\right]\gamma}I_{\mu N_t N_r - 0.5}\left(\left[\frac{2\mu H}{\tilde{\gamma}}\right]\gamma\right), (3)$$

which, for compactness, is written as:

$$f_\gamma(\gamma) = \psi\gamma^{m-1}e^{-\beta\gamma}I_v(\xi\gamma), \quad (4)$$

where $\psi = \frac{2\sqrt{\pi}h^{\mu N_t N_r}}{\Gamma(\mu N_t N_r)H^{m-1}}\left(\frac{\mu}{\tilde{\gamma}}\right)^m$, $m = \mu N_t N_r + 0.5$, $\beta = \left[\frac{2\mu h}{\tilde{\gamma}}\right]$, $\xi = \left[\frac{2\mu H}{\tilde{\gamma}}\right]$ and $v = m - 1$.

Similarly, it was shown in [24] that the sum of $L$ i.i.d. generalized $\kappa - \mu$ shadowed random variables (R.V.) with parameters $\kappa$, $\mu$, $m$ and $\tilde{\gamma}$ is also another $\kappa - \mu$ shadowed R.V. with the parameters $\kappa$, $L\mu$, $Lm$ and $L\tilde{\gamma}$. Following this result, the received power PDF with $N_t$ transmit antenna and $N_r$ received antenna can be expressed as:

$$f_\gamma(\gamma) = \frac{\tilde{\mu}^{\tilde{\mu}}\tilde{m}^{\tilde{m}}(1+\kappa)^{\tilde{\mu}}}{\Gamma(\tilde{\mu})(\tilde{\mu}\kappa+\tilde{m})^{\tilde{m}}\eta^{\tilde{\mu}}}\gamma^{\tilde{\mu}-1}e^{-\left[\frac{\tilde{\mu}(1+\kappa)}{\eta}\right]\gamma} {}_1F_1\left(\tilde{m};\tilde{\mu};\left[\frac{\tilde{\mu}^2\kappa(1+\kappa)}{[\tilde{\mu}\kappa+\tilde{m}]\eta}\right]\gamma\right), (5)$$

where $\tilde{\mu} = N_t N_r \mu$, $\tilde{m} = N_t N_r m$, $\eta = N_t N_r \tilde{\gamma}$, which is written (for compactness) in the form:

$$f_\gamma(\gamma) = \psi\gamma^{\tilde{\mu}-1}e^{-\beta\gamma} {}_1F_1(\tilde{m};\tilde{\mu};\zeta\gamma), \quad (6)$$

where $\psi = \frac{\tilde{\mu}^{\tilde{\mu}} \tilde{m}^{\tilde{m}} (1+\kappa)^{\tilde{\mu}}}{\Gamma(\tilde{\mu})(\tilde{\mu}\kappa+\tilde{m})^{\tilde{m}} \eta^{\tilde{\mu}}}$, $\beta = \left[\frac{\tilde{\mu}(1+\kappa)}{\eta}\right]$, and $\zeta = \left[\frac{\tilde{\mu}^2 \kappa (1+\kappa)}{[\tilde{\mu}\kappa+\tilde{m}]\eta}\right]$.

The expressions in (4) and (6) will be used later in our unified bit error rates analysis.

### D. The AWGGN Environment

For simplicity, many wireless communications researchers have considered the additive Gaussian model as their noise conditions, written in terms of the well-known Gaussian $Q$-function. Wireless communications channels might, however, be subjected to different sources of noise that is non-Gaussian. Examples of the non-Gaussian noise include the Laplacian and the Gamma noise (see [1] [26] [27] for more discussion on the different noise environments). The AWGGN is a relatively new and generalized noise model that includes many of well-known noise models as special cases. It is represented written analytically using the generalized $Q$–function:

$$Q_a(x) = \frac{a \Lambda_0^{2/a}}{2\Gamma(1/a)} \int_x^\infty e^{-\Lambda_0^a |u|^a} du = \frac{\Lambda_0^{2/a-1}}{2\Gamma(1/a)} \Gamma(1/a, \Lambda_0^a |x|^a). \quad (7)$$

where $\Lambda_0 = \sqrt{\Gamma(3/a)/\Gamma(1/a)}$, $\Gamma(\cdot)$ denotes the famous gamma function, and $a$ being the noise parameter. Table III shows a summary of the special cases of (7).

TABLE III: Special Cases of AWGGN.

| Noise Dist. | Impulsive | Gamma | Laplacian | Gaussian | Uniform |
|---|---|---|---|---|---|
| $a$ | 0.0 | 0.5 | 1.0 | 2.0 | $\infty$ |

As reported in [1], the expression in (7) can be accurately well-approximated (the robustness of the approximation is shown in [1]) as a simple sum of four scaled and decaying exponential functions, using the curve-fitting model (and not a truncated-series) that is given as:

$$Q_a(\sqrt{x}) \approx \sum_{i=1}^{4} p_i e^{-q_i x}, \quad (8)$$

where the fitting parameters, $p_i$ and $q_i$ are evaluated as given and discussed in [1] and using Levenberg-Marquardt nonlinear curve fitting, with sample fitting values for different cases of the noise parameter $a$ being presented in Table IV.

TABLE IV: Fitting Parameters of $Q_a(\sqrt{\cdot})$ Approximation

| $a$ | $p_1$ | $p_2$ | $p_3$ | $p_4$ | $q_1$ | $q_2$ | $q_3$ | $q_4$ |
|---|---|---|---|---|---|---|---|---|
| 0.5 | 44.920 | 126.460 | 389.400 | 96.540 | 0.130 | 2.311 | 12.52 | 0.629 |
| 1 | 0.068 | 0.202 | 0.182 | 0.255 | 0.217 | 2.185 | 0.657 | 12.640 |
| 1.5 | 0.065 | 0.149 | 0.136 | 0.125 | 0.341 | 0.712 | 10.57 | 1.945 |
| 2 | 0.099 | 0.157 | 0.124 | 0.119 | 1.981 | 0.534 | 0.852 | 10.268 |
| 2.5 | 0.126 | 1.104 | -1.125 | 0.442 | 9.395 | 0.833 | 0.994 | 1.292 |

The relative absolute error of this approximation model was discussed in [1], and using only for exponential functions in the model provided a compromise between the computational complexity and non-distinguishable results from the exact evaluation, as will be seen in section IV. This approximation will be used in section III in our unified performance expression.

### III. The Unified Bit Error Rates Analysis

#### A. Average Bit Error Rate

The average bit error rate (ABER) due to a wireless fading channel can be analytically found by averaging the bit error rate of the noisy channel over the PDF of the instantaneous signal-to-noise ratio (SNR), $f_\gamma(\gamma)$. With the AWGGN noise assumption, the averaging process can be expressed as:

$$P_e = \mathcal{A} \int_0^\infty f_\gamma(\gamma) Q_a(\sqrt{\mathcal{B}\gamma}) d\gamma, \quad (9)$$

where $Q_a(\sqrt{\cdot})$ models conditional error probability given the channel state in an AWGGN, and $\mathcal{A}$ and $\mathcal{B}$ are modulation-scheme dependent, as shown in Table V.

Using (9), we will derive the ABER expressions for the $\eta - \mu$ and $\kappa - \mu$ shadowed generalized fading models subjected to AWGGN next.

TABLE V: $\mathcal{A}$ and $\mathcal{B}$ Values For Different Modulations

| Modulation Scheme | Average SER | $\mathcal{A}$ | $\mathcal{B}$ |
|---|---|---|---|
| BFSK | $= Q_a(\sqrt{\gamma})$ | 1 | 1 |
| BPSK | $= Q_a(\sqrt{2\gamma})$ | 1 | 2 |
| QPSK, 4-QAM | $\approx 2Q_a(\sqrt{\gamma})$ | 2 | 1 |
| M-PAM | $\approx \frac{2(M-1)}{M} Q_a\left(\sqrt{\frac{6}{M^2-1}\gamma}\right)$ | $\frac{2(M-1)}{M}$ | $\frac{6}{M^2-1}$ |
| M-PSK | $\approx 2Q_a\left(\sqrt{2\sin^2\left(\frac{\pi}{M}\right)\gamma}\right)$ | 2 | $2\sin^2\left(\frac{\pi}{M}\right)$ |
| Rectangular M-QAM | $\approx \frac{4(\sqrt{M}-1)}{\sqrt{M}} Q_a\left(\sqrt{\frac{3}{M-1}\gamma}\right)$ | $\frac{4(\sqrt{M}-1)}{\sqrt{M}}$ | $\frac{3}{M-1}$ |
| Non-Rectangular M-QAM | $\approx 4 Q_a\left(\sqrt{\frac{3}{M-1}\gamma}\right)$ | 4 | $\frac{3}{M-1}$ |

#### B. $\eta - \mu$ fading

Substituting (4) into (9), then (9) can be re-written as:

$$P_e = \mathcal{A}\psi \int_0^\infty \gamma^{m-1} e^{-\beta\gamma} I_v(\xi\gamma) Q_a(\sqrt{\mathcal{B}\gamma}) d\gamma, \quad (10)$$

and by utilizing the generalized $Q_a(\sqrt{\cdot})$ approximation in (8), then (10) can then be written in the form:

$$P_e = \sum_{i=1}^{4} [\mathcal{A}\psi p_i] \int_0^\infty \gamma^{m-1} e^{-\tilde{\beta}\gamma} I_v(\xi\gamma) d\gamma, \quad (11)$$

which can be evaluated in a closed-form [28], evaluating to:

$$P_e = \sum_{i=1}^{4} \Psi \, {}_2F_1\left(\left[\frac{m+v}{2}, \frac{m+v+1}{2}\right]; [1+v], \frac{\xi^2}{\tilde{\beta}^2}\right), \quad (12)$$

where $\tilde{\beta} = [\beta + q_i \mathcal{B}]$, $\Psi = \frac{[\mathcal{A}\psi p_i] \xi^v \Gamma(m+v)}{2^v \tilde{\beta}^{m+v} \Gamma(v+1)}$, and ${}_2F_1([\cdot,\cdot];\cdot;\cdot)$ is the Gauss hypergeometric function [20].

This new derived expression in (12) is simple, generic and generalized for the ABER analysis in the $\eta - \mu$ two formats deploying STBC reception in AWGGN, and applies directly to all the special cases of the $\eta - \mu$ fading model.

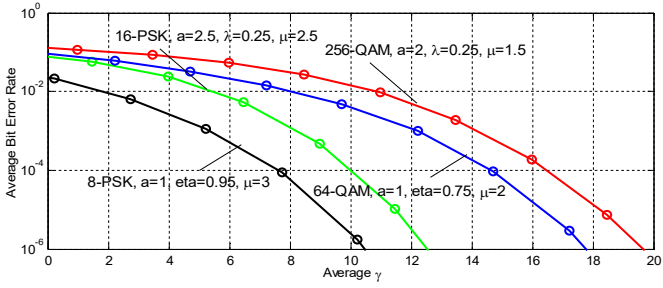

Fig. 1: ABER for different modulation schemes in generalized $\eta - \mu$ fading subjected to different types of noise ($N_t = N_r = 2$).

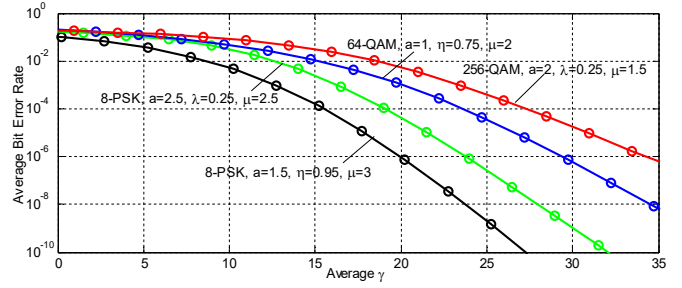

Fig. 2: ABER for different modulation schemes in generalized $\eta - \mu$ fading subjected to different types of noise ($N_t = N_r = 1$).

### C. $\kappa - \mu$ shadowed fading

Substituting (6) into (9), then (9) can be rewritten as:

$$P_e = \mathcal{A}\psi \int_0^\infty \gamma^{\tilde{\mu}-1} e^{-\beta\gamma} \, _1F_1(\tilde{m}; \tilde{\mu}; \zeta\gamma) \, Q_a(\sqrt{\mathcal{B}\gamma}) d\gamma, \quad (13)$$

and using the generalized $Q_a(\sqrt{\cdot})$ approximation from (8), one arrives at the integral:

$$P_e = \mathcal{A}\psi \sum_{i=1}^{4} p_i \int_0^\infty \gamma^{\tilde{\mu}-1} e^{-[\beta+q_i\mathcal{B}]\gamma} \, _1F_1(\tilde{m}; \tilde{\mu}; \zeta\gamma) \, d\gamma. \quad (14)$$

With simple variable transform, as $z = \zeta\gamma$, (15) is evaluated in a simple closed form as:

$$P_e = \sum_{i=1}^{4} \Psi \, _2F_1([\tilde{m}, \tilde{\mu}]; [\tilde{\mu}]; \tilde{\beta}^{-1}), \quad (15)$$

with $\tilde{\beta} = [\beta + q_i\mathcal{B}]/\zeta$ and $\Psi = [\mathcal{A}\psi p_i \Gamma(\tilde{\mu})]/[\zeta^{\tilde{\mu}} \tilde{\beta}^{\tilde{\mu}}]$.

This expression is new, and it generalizes the results to all the special cases of the shadowed $\kappa - \mu$ fading and AWGGN special cases, as given in tables II and III, respectively, while deploying MIMO STBC. Note that (12) and (15) are two different ways to calculate ABER in $\eta - \mu$ fading, since it is a special case of the $\kappa - \mu$ shadowed fading [17].

## IV. NUMERICAL RESULTS

This section illustrates sample ABER analytical results for the derived expressions in (12) and (15), which are compared with the numerically obtained results. Five test cases are presented that targets the analysis of the effect of the fading parameters and the additive noise environment, given different modulation schemes. For the purpose of clarity of the plots in the figures, the use of legend is avoided, however, please note that solid curves represent the numerical results whereas the overlaid patterns represent the derived analytical expressions. Moreover, please note that the fading and noise conditions as well as modulation schemes and constellation orders are also indicated on the plots.

In the first and second tests, we consider the $\eta - \mu$ fading with $N_t = N_r = 2$ and $N_t = N_r = 1$, respectively. The generated results are shown in Fig. 1 and 2, respectively. One can clearly see the excellent match between the numerical results and those obtained using (12). Moreover, as one expects, doubling the number of antennas at the transmitter's and receiver's sides in Fig. 1 has significantly improved the ABER performance as compared to Fig. 2. Furthermore, the effect of higher $\mu$ values, the number of multipath clusters, similarly improves the ABER performance significantly, in contrast to the effect of $\eta$ values that has minor effects on the ABER performance. Another observation from these figures is that the effect of the noise parameter is evident and is not negligible, where higher $a$ shows better performance. This suggests that a careful study of the noise in a given system deployment environment is very important for an accurate system modeling and analysis, which supports earlier reported results from earlier studies on the non-additive white Gaussian noise literature.

Similarly, the third and fourth test scenarios follow the same assumption of the first two tests, but assuming $\kappa - \mu$ shadowed fading with the parameters shown in the plots. The generated results, shown in Fig. 3 and Fig. 4 indicate a very good match between (15) and the numerically obtained results. As expected, the performance gains due to the utilization of the multiple transmit and receive antenna can be clearly observed by comparing Fig. 3 vs. Fig. 4. Moreover, one can draw similar conclusions on the effect of the number of multipath clusters, $\mu$, and the noise environment represented by the parameter $a$. The plots also suggest that a higher value of $\kappa$, which represents the ratio of the dominant LOS component and the total power of the scattered waves, the system's ABER performance is expected to be better. This is a very important result especially for wireless systems that are designed to operate in LOS conditions. The shadowing parameter $m$, in stark contrast to $\kappa$, has a negative effect on the ABER performance. That is, the more severe the signal's shadowing, the more the ABER performance degrades as can be intuitively concluded.

Finally, as a further verification step, we regenerate in Fig. 5 and Fig. 6 the plots given in [19], using our new ABER expression (15) for the $\kappa - \mu$ shadowed fading channels by selecting the corresponding parameter values from Table II, while $N_t = N_r = 1$ and $a = 2$ (additive Gaussian noise). We emphasize here that, as was shown in [17], the $\kappa - \mu$ shadowed fading model encloses the $\eta - \mu$ model as a special case. By comparing Fig. 5 and Fig. 6 against [17], one can see indistinguishable ABER curves against those presented in Natalia's work in [17]. The plots were similarly verified by using the results from [19], and in the case of the $\eta - \mu$ fading, the curves in Fig. 6 were also compared against those obtained from (12) but skipped here to avoid redundancy.

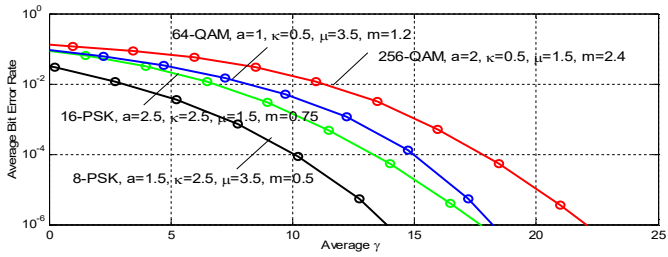

Fig. 3: ABER for different modulation schemes in generalized $\kappa - \mu$ shadowed fading subjected to different types of noise ($N_t = N_r = 2$).

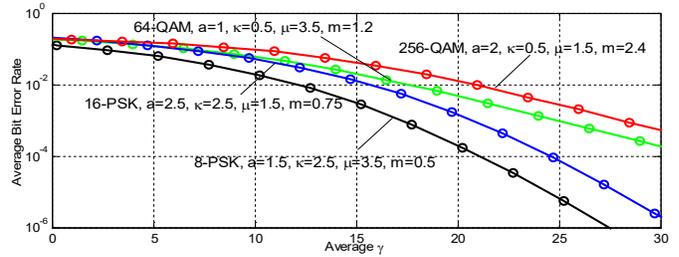

Fig. 4: ABER for different modulation schemes in generalized $\kappa - \mu$ shadowed fading subjected to different types of noise ($N_t = N_r = 1$).

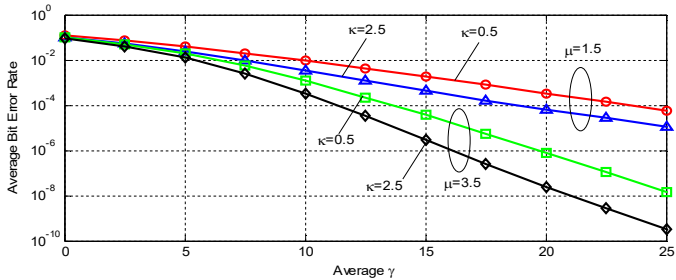

Fig. 5: Regenerated figure from ( [19] Fig. 1).

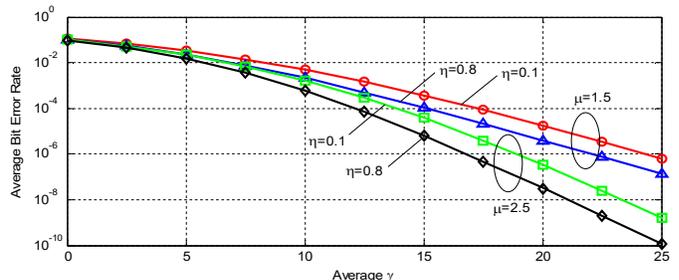

Fig. 6: Regenerated figure from ( [19] Fig. 2).

It is concluded from this section and the studied performance tests that the new derived unified expressions are very accurate to study STBC in generalized $\eta - \mu$ and $\kappa - \mu$ fading channels.

## V. CONCLUSION AND FUTURE WORK

This paper presented novel unified performance analysis of STBC in MIMO network. The new derived expressions for the ABER are unified, simple and applicable for all coherent modulation schemes assuming $\eta - \mu$ and shadowed $\kappa - \mu$ fading channels and the AWGGN environment. The result obviates the need to re-evaluate similar expressions for many fading and noise models in a piecemeal fashion. The presented intensive testing proves the accuracy of our expressions and their suitability to study STBC transmission in the aforementioned fading and noise conditions and the effect of the fading and noise parameters on such wireless communications scenarios.

For a future work, we plan to utilize the measurements used in [29] [30] for channel modeling using the $\eta$-$\mu$ and $\kappa$-$\mu$ shadowed fading conditions, as well as to study the performance of [31] [32] [33] [34] [35] in these fading generalized channels.